**Research Article**

# Implementation of a Radiation Protection System at Four Hospitals in Ethiopia

**Gebre Mesay Geletu\*, Fikru Abiko, and Shamble Sahlu**

*Department of Physics, Wolkite University, Ethiopia*

## Abstract

Diagnostic imaging procedure using ionizing radiation has been growing due to its enormous benefits. However, using ionizing radiation is also associated with harmful biological effects. Therefore it is important to take consideration on harmful effects of ionizing radiation and have adequate awareness on the area, in order to protect people without limiting the benefits. In Ethiopia, the status of radiation protection practices have not been studied previously, which remain nearly unknown especially around the study area. Thus, in filing this gap the present study initiated to assess status of implementation of radiation protection system. General observation, questionnaire and dose rate measurements were used to get all necessary data for the study. The dose rate measurements at control room (CR), waiting room (WR) and inside X-ray room (XR) were carried out using well calibrated Thermo FH 40 G-L10 survey meter. The results of the general observation and the questionnaire showed poor practice of radiation protection system. The dose rate recordings from CR and WR, when the X-ray machine being switched on or off are within permissible allowed value at all hospitals. XR machine off measurements at two locations are also within the range allowed value. However, XR machine on recordings shows a high level of radiation. This causes very high level unnecessary exposure situation for helper as a result of poor practices of radiation protection system which intern extremely increases the population dose. Therefore there is a great need of radiation protection agency activity at all Hospitals. Considering this scenario we suggest continuous training programs for radiographers; all Hospitals must full fill more modern radiation protection monitoring equipment and should have radiation protection advisers to monitor the radiation protection practices.

## ABBREVIATIONS

CR: Control Room; WR: Waiting Room; XR: X-Ray Room; XR1: X-Ray Room Location One; XR2: X-Ray Room Location Two

## INTRODUCTION

Recently, because of its tremendous benefits, medical imaging procedures involving the use of ionizing radiation is growing very rapidly all over the world. However, using ionizing radiation is also associated with risks due to its harmful biological effects upon human exposure. There are two main types of undesirable biological effects, such as stochastic and or deterministic effects [1,2]. Deterministic effect has a threshold of dose value and severity of the effect increases with dose whereas stochastic effects have no dose threshold and the probability of an effect increases with dose. Therefore it is very essential to take consideration on harmful effect of ionizing radiation and have adequate awareness on the area [3]. Considering the harmful biological effects of ionizing radiation, there has been radiation protection practice for monitoring and assessing the levels of exposure and keep one's exposure to ionizing radiation As Low As Reasonably Achievable (ALARA). As stated in ICRP publication 105 the main aim of radiation protection is to provide an appropriate standard of protection for people and the environment without unduly limiting the beneficial practices giving rise to radiation exposure. In Ethiopia, though diagnostic imaging facilities have been growing in government and private Hospitals, the availability and quality of radiological service are still poor. This can be justified with the survey carried out by Shimelis D. et al. [5], in Addis Ababa public hospitals. Radiation safety for patients, staffs and public around are inadequate level. Moreover, the present status of implementation of radiation protection system has not been studied previously, which remain unknown especially around remote areas. Thus the main objectives of this study are to investigate the potential radiation sources and type of radiation, and the status of radiation protection practice at four Hospitals near to Wolkite city.

## MATERIALS AND METHODS

The study area covers all four Hospitals found near to Wolkite town. Woliso Hospital, which is found in Oromiya Region in Woliso city. Butajera and Mercy Hospitals found in SNNP region in Gurage Zone in Botajira city and 18 km from Butajira city respectively [6]. Atat Hospital, is about 20 km from Wolkite town.

In order to collect both qualitative and quantitative data the study was divided into three separate works: (I) **Observation:-** this stage is used in identification of radiation sources including type of radiation; and to inspect implementation of the radiation protective devices and circumstances under which people are exposed. (II) **Questionnaire: -** The questionnaire consists of twenty-two questions and it is filled by radiographers. The purpose of the questionnaire based survey include: to get information about the general understanding of harmful effects of ionizing radiation, and radiation protection guide lines; , and to assess frequent radiation protection training. (III) **Measurement: -** The last stage of the data collection was measurement of the radiation dose rate when the identified radiation source (Diagnostic X-ray unit) is on and off. Well calibrated Thermo FH 40 G-L10 Survey Meter was used. The measurements were taken in three different rooms such as inside x-ray room (XR), control room (CR) and outside X-ray room near to main entrance which is waiting area (WR). CR and WA measurements were taken at single location near to the door. Whereas, XR dose rate measurements were taken at two locations: along a side of the vertical wall chest stand (named as XR1) and along a side to the coach or table (named as XR2) see Figure 1. The measurement procedure is as follows: First at all locations the background radiation dose rate measurements were taken before the X-ray machine was switched on; and secondly, when X-ray machine is on, the corresponding dose rate was recorded. When the machine is on, the most common exposure parameters for each Hospital were set see Table 1.

## RESULTS AND DISCUSSION

As seen during the visit for observation all hospitals provide the basic level of diagnostic radiology services. Information about

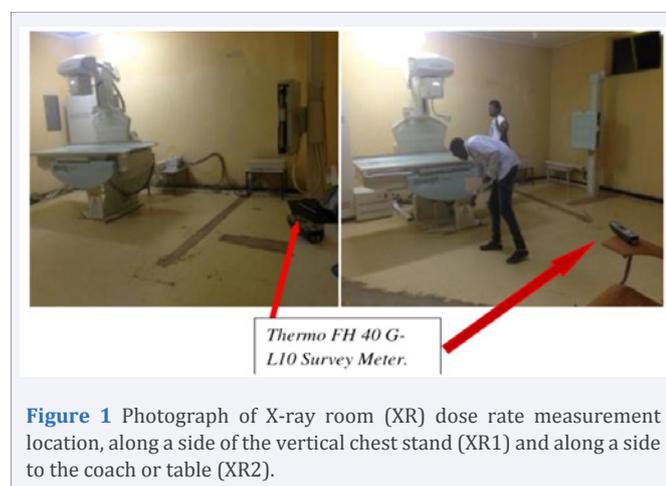

*Thermo FH 40 G-L10 Survey Meter.*

**Figure 1** Photograph of X-ray room (XR) dose rate measurement location, along a side of the vertical chest stand (XR1) and along a side to the coach or table (XR2).

**Table 1:** Exposure parameters for X-ray machine on measurements.

| Name of Hospital | KVp | mAs | Exposure time (se) | FFD |
|---|---|---|---|---|
| Atat | 75 | 100 x 0.4 | 0.4 | 1.5 |
| Butajira | 70 | 250 x 0.08 | 0.08 | 1.5 |
| Mercy | 58 | 250 x 0.09 | 0.09 | 1.5 |
| Woliso | 70 | 160 x 0.2 | 0.2 | 1.5 |

**Abbreviations:** KVP: Kilovoltage Peak; MAS: Milliam per second; FFD: Focus Film Distance

diagnostic X-ray machines available at all Hospitals are given in Table 2. Only Butajira Hospital has digital radiography the other three have plane film radiography.

Though most radiation shielding devises are available see Table 3, it was observed that radiographers, patients, and helpers do not use these devises during X-ray examination. Neck collar and lead cap are not available at all Hospitals. Dose limits for staff and public were published by the International Commission on Radiological Protection (ICRP) in Publication 103. Thus, it is very important to estimate the dose, for the purpose of radioprotection procedures and restrict the hazards to human health. However, during the study period it was found that all radiographers at all hospitals are not using personal dose monitoring badge. There is no written guideline and directions at all Hospitals for proper implementation of radiation protection system. Ionizing radiations regulations 1999, Regulation 14 sates: "all employees involved in the work with ionizing radiation, including management, needs training to help develop and sustain a commitment to restricting exposure wherever this is reasonably practicable. The employer usually needs to provide training to ensure employees are competent where a system of work or personal protective equipment is provided to restrict exposure." But, there is no radiation protection short term trainings and courses in all Hospitals for radiographer and related workers. Thus all results provide evidence for the existing poor practice of radiation protection system.

The measurement results of radiation dose rate level at WR, CR and XR when the X-ray machine is on and off are presented in Table 4 and 5. As shown in Table 4, CR machine on recordings of the mean dose rate ranged from lowest value 0.033 μSv/hr to highest value 0.158 μSv/hr. While machine off CR measurements, ranges from lowest 0.017 μSv/hr to 0.15 μSv/hr highest. Comparing machine on maximum dose rate recording 0.158 μSv/hr with the maximum machine off dose rate 0.15 μSv/hr, the resulting difference is 0.008 μSv/hr which is insignificant. Similarly WR machine on measurement of the mean dose rate ranged from minimum value 0.08 μSv/hr to 0.15 μSv/hr maximum value. While machine off WR measurements ranged from minimum value 0.1 μSv/hr to 0.136 μSv/hr maximum. The maximum mean dose rate recorded when machine is on 0.15 μSv/hr comparing with maximum background mean dose rate 0.136 μSv/hr, the difference is only 0.014 μSv/hr. This suggests that when the X-ray machine on there no additional radiation coming to CR and WR in all Hospitals, which is good in view of radiation protection.

Regarding XR results, as presented in the Table 5 machine off minimum mean dose rate, at XR1 is 0.12 μSv/hr while the maximum mean dose rate, at XR1, is 0.14 μSv/hr. Similarly machine off minimum mean dose rate, at XR2 is 0.12 μSv/hr while the maximum mean dose rate, at XR2, is 0.148 μSv/hr. These results are in agreement with CR and WR similar machine off measurements. The background radiation (machine off) is consistently lower than the corresponding value for the machine on measurements. This indicates the higher level of ionizing radiation in XR when the machine is on than when the machine off in contrast to CR and WR. Considering only XR1 dose rate readings, the results ranged from 40.86 ±1.33 μSv/hr obtained in Woliso Hospital to 142 ± 2.16 μSv/hr for Mercy Hospital.

**Table 2:** Diagnostic X-ray machine information.

| Name of Hospital | Type of institutions | X-ray Machine Information | | | |
| | | Machine Manufacturer | Digital/plane Film radiography | Tube model | Tube Manufacturer |
|---|---|---|---|---|---|
| Atat | NGO | Wandong | Plane film | xd51-20.40/100-T1A | Kee Hing Cheung Kee (KHCK) |
| Butajira | Government | Shimadzu | Digital | G-292 | Varex Imaging |
| Mercy | NGO | Shimadzu | Plane film | Circlex ½ P13Dk 805 | Shimadzu Kee Hing |
| Woliso | NGO | Wandong | plane film | XD51-20.40/100 T1A | Cheung Kee (KHCK) |

**Abbreviations:** NGO: Non-governmental organization

**Table 3:** Availability of radiation protection shielding materials.

| Name of hospital | Lead apron | | Gonad shields | | Lead gloves | | lead Glass goggle | | lead cap | neck collar |
| | Available | No one using | Available | No one using | Available | No one using | Available | No one using | Available | Available |
|---|---|---|---|---|---|---|---|---|---|---|
| Mercy | ✓ | ✓ | ✓ | ✓ | ✓ | ✓ | ✓ | ✓ | ✗ | ✗ |
| Butajira | ✓ | ✓ | ✓ | ✓ | ✓ | ✓ | ✓ | ✓ | ✗ | ✗ |
| Atat | ✓ | ✓ | ✓ | ✓ | ✓ | ✓ | ✓ | ✓ | ✗ | ✗ |
| Weliso | ✓ | ✓ | ✓ | ✓ | ✓ | ✓ | ✓ | ✓ | ✗ | ✗ |

**Table 4:** Control and waiting rooms mean dose rate measurement result.

| Name of Hospitals | Number of Measurements | Control room on Machine (µSv/hr) | Machine off (µSv/hr) | Waiting room | |
| | | | | Machine on (µSv/hr) | Machine off (µSv/hr) |
|---|---|---|---|---|---|
| Atat | 4 | 0.158 | 0.117 | 0.08 | 0.13 |
| Butajira | 4 | 0.07 | 0.13 | 0.15 | 0.12 |
| Mercy | 4 | 0.033 | 0.017 | 0.1 | 0.1 |
| Woliso | 4 | 0.1 | 0.15 | 0.141 | 0.136 |

**Table 5:** X-ray room mean dose rate measurement result, Values in brackets indicate range.

| Number of hospitals | Place | Number of measurements | Machine off mean dose ±SD ( µSv/hr) | | Machine on mean dose ±SD ( µSv/hr) |
|---|---|---|---|---|---|
| Atat | XR 1 | 4 | 0.14±0.02; | (0.122-0.176) | 41.47±13.26; (31.4-60.2) |
| | XR2 | 4 | 0.148±0.03; | (0.13-0.85) | 36.066±0.52; (35.7-36.8) |
| Butajira | XR1 | 4 | 0.12±0.02; | (0.105-0.139) | 50.83±8.25; (40-60) |
| | XR2 | 4 | 0.14±0.02; | (0.114-0.172) | 26.33± 3.52; (23.6-31.3) |
| Mercy | XR1 | 4 | 0.14±0.01; | (0.13-0.15) | 142±2.16; (140-145) |
| | XR2 | 4 | 0.12±0.017; | (0.10-0.14) | 52.83±3.26; (50-57.4) |
| Woliso | XR1 | 4 | 0.132±0.01; | (0.122-0.147) | 40.8633±1.33; (39.5942.7) |
| | XR2 | 4 | 0.136±0.01; | (0.125-0.13 ) | 26.33± 1.99; (23.6-28.3) |

**Abbreviations:** XR1: X-ray room location one; X X-ray room location two

According to the results there is a variation from Hospitals to Hospitals. Similarly XR2 measurement ranged from minimum value 26.33 µ Sv/hr at Woliso and Botajira Hospitals 52.83 ± 3.26 µSv/hr maximum Mercy Hospital. At all Hospitals XR1 results are greater than XR2 results. This can reveal the variation of the dose rate level from point to other point in the room.

## CONCLUSION

Generally the results of the study shows that in all centers application of shielding devices are mostly neglected and dose monitoring for radiographers and patients are a big problems. Furthermore, there is no training for radiographers to update themselves though they worked for years, this resulted bad working practice of radiographers to protect themselves, patients, staff and the members of the public against unnecessary exposure. Based on the dose rate results there is no leakage to WR and CR in all Hospitals, indicating the good situation of the structural radiation protection. However, within the X-ray rooms doses did enhance with the machine being switched on (as everyone would have expected) but the people do not wear their personal radiation protection equipment. This resulted very bad situation in which enormously increases the population risk of radiation exposure due to the ignorance behavior or missing legislation. Therefore there is a great need for rules, regulation and radiation protection agency activity in all Hospitals. Based on the overall results we suggest continuous training programs; all Hospitals must full fill more modern radiation protection monitoring equipment and should have a radiation protection legislations and advisers to monitor the radiation protection practices, so that any undesirable effect can be minimized and to ensure compliance with radiation safety regulations.

## ACKNOWLEDGEMENTS

The authors are grateful to Wolkite University for funding this research. We thank Ethiopian Radiation Protection Authority (ERPA) and Mr. Nurdin for cooperating us in this study. Our acknowledgment is extended to the managers of four hospitals and radiographers.